\documentclass[amsmath, amssymb, aps, twocolumn]{revtex4-1}
\usepackage{graphicx}
\usepackage{amsmath}
\usepackage{bm}
\usepackage{mathrsfs}
\usepackage{booktabs}

\begin{document}

\title{Precise determination of two-carrier transport properties in the topological insulator TlBiSe$_2$}

\author{*G. Eguchi$^1$}
\email{geguchi@kuee.kyoto-u.ac.jp}
\author{K. Kuroda$^2$} 
\altaffiliation[Current address: ]{Institute for Solid State Physics, the University of Tokyo, Japan}
\author{K. Shirai$^2$}
\author{Y. Ando$^1$}
\author{T. Shinjo$^1$}
\author{A. Kimura$^2$} 
\author{M. Shiraishi$^1$}

\affiliation{$^1$Department of Electronic Science and Engineering, Graduate School of Engineering, Kyoto University, Kyoto 615-8510, Japan}
\affiliation{$^2$Graduate School of Science, Hiroshima University, Kagamiyama, Higashi-Hiroshima 739-8526, Japan}

\date{\today}

\begin{abstract}
We report the electric transport study of the three-dimensional topological insulator TlBiSe$_2$. We applied a newly developed analysis procedure and precisely determined two-carrier transport properties. Magnetotransport properties revealed a multicarrier conduction of high- and low-mobility electrons in the bulk, which was in qualitative agreement with angle-resolved photoemission results~[K. Kuroda $et~al.$, Phys. Rev. Lett. $\bm{105}$, 146801 (2010)]. The temperature dependence of the Hall mobility was explained well with the conventional Bloch-Gr{\"u}neisen formula and yielded the Debye temperature $\varTheta_{\rm{D}}=113 \pm 14$~K. The results indicate that the scattering of bulk electrons is dominated by acoustic phonons.

\end{abstract}

\maketitle

\section{Introduction}
Electric and spin transport in three-dimensional (3D) topological insulators has received a great deal of attention because of the insulators' surface Dirac fermions with the unique spin texture. The presence of the spin texture has been confirmed by several experiments employing angle-resolved photoemission spectroscopy (ARPES), and a variety of novel transport phenomena have been proposed~\cite{RevModPhys.82.3045,RevModPhys.83.1057,doi:10.7566/JPSJ.82.102001,PhysRevLett.106.257004,NatNano.9.218,doi:10.1021/nl5026198,doi:10.1021/nl502546c,RevModPhys.82.3045,RevModPhys.83.1057,PhysRevB.82.155457}. In real materials, however, unavoidable bulk conduction often hinders surface conduction.

Electric transport in 3D topological insulators has been analyzed with the so-called two-carrier model; i.e., the parallel conduction of carriers with different mobility~\cite{PhysRevB.82.241306,PhysRevLett.107.016801,PhysRevLett.109.116804}. Within the model, the resistivity $\rho_{xx}$ in a magnetic field $B=\mu_0 H$ is expressed by the Equation
\begin{equation}
\rho_{xx}(B)= \frac{n_1q_1\mu_1 + n_2q_2\mu_2 + (n_1q_1\mu_2+n_2q_2\mu_1)\mu_1\mu_2B^2} {(n_1q_1\mu_1+n_2q_2\mu_2)^2 + (n_1q_1+n_2q_2)^2\mu_1^2\mu_2^2B^2},
\label{eq1}
\end{equation}
where $q_1$ and $q_2$, $n_1$ and $n_2$, and $\mu_1$ and $\mu_2$ are respectively the charges, densities and mobilities of the carriers. Despite its usability, the model contains four active parameters, which lead to large errors in most cases. The errors have not been discussed and their statistical reliability remains unclear. This is a fundamental issue requiring a more elaborate analysis procedure. 

In this article, we report the electric transport of the 3D topological insulator TlBiSe$_2$. The  material is known as one of the simplest and most practical materials for the transport studies~\cite{0295-5075-90-3-37002,PhysRevLett.105.036404,PhysRevLett.105.136802,PhysRevLett.105.146801,PhysRevB.91.205306,PhysRevB.90.201307}. Two-carrier transport properties were precisely determined by applying a newly developed analysis procedure. The procedure enables an analysis with two active parameters and with sufficiently small errors. The magnetotransport properties were well explained accounting for high- and low-mobility electrons in the whole temperature range. The carrier densities were $(0.61 \pm 0.01) \times 10^{19}$~/cm$^3$ and $(4.45 \pm 0.01) \times 10^{19}$~/cm$^3$, which were $10^5$ times the values expected from the surface. These results indicate that multicarrier conduction originated from the bulk electrons, and the scattering was insensitive to the magnetic field. The temperature dependence of the Hall mobility exhibited a metallic behavior over the whole temperature range, and was explained well by the Bloch-Gr{\"u}neisen formula. This indicates that the scattering of the bulk electrons was dominated by acoustic phonons. It should be noted that analyses proposed in this article allow a precise identification of the two-carrier transport properties and are therefore useful for all other multicarrier systems.

\section{Experimental}
\begin{figure}[h]
\begin{center}
\includegraphics[width=\columnwidth,clip]{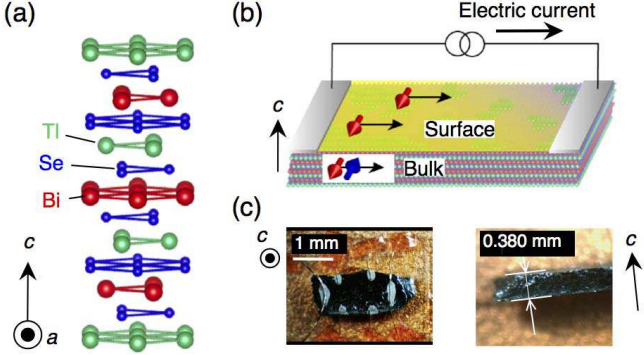}
\caption{(Color online) (a)~A schematic of the TlBiSe$_2$ crystal with hexagonal coordinates~(space group: $R\bar{3}m$~\cite{PhysRevLett.105.036404}). (b)~Schematics of parallel conduction of the spin-polarized surface and bulk. The cleaved surface forms thallium islands on the selenium layer~\cite{PhysRevB.88.245308}. (c)~Optical images of the top surface and side surface. The sample thickness was $d=0.380$~mm.}
\label{fig1}
\end{center}
\end{figure}

The TlBiSe$_2$ sample used in this study was taken from the same batch of Ref.~\cite{PhysRevB.91.205306}.  As reported in Ref.~\cite{PhysRevB.91.205306}, a significant variation of the chemical composition was revealed between the starting substances and synthesized crystals; the variation resulted in the spontaneous electron doping. The sample, which was synthesized from the starting composition ${\rm{Tl}}:{\rm{Bi}}:{\rm{Se}} = 1:1:2$ in stoichiometric ratio, corresponded to Tl$_{1-x}$Bi$_{1+x}$Se$_{2-\delta}$ ($x=0.064, \delta=0.29$)~\cite{PhysRevB.91.205306,comment1}. 
Electric transport experiments were performed using the conventional six-probe technique down to 2~K with a commercial apparatus (Quantum Design PPMS). Electric contacts were made onto the cleaved surface with room-temperature-cured silver paste under atmospheric conditions. Geometries including the distances of electrodes and the thickness of the sample were determined from optical microscope images. Schematics of the crystal and parallel conduction of the surface and bulk are presented in Figure~\ref{fig1}(a), (b). The optical microscope images are also shown in Figure~\ref{fig1}(c). The sample had a thickness of 0.380~mm. Magnetic fields (maximum strength of $\pm9$~T) were applied parallel to the $c$-axis.

\section{Results and Analyses}
\begin{figure}[h]
\begin{center}
\includegraphics[width=\columnwidth,clip]{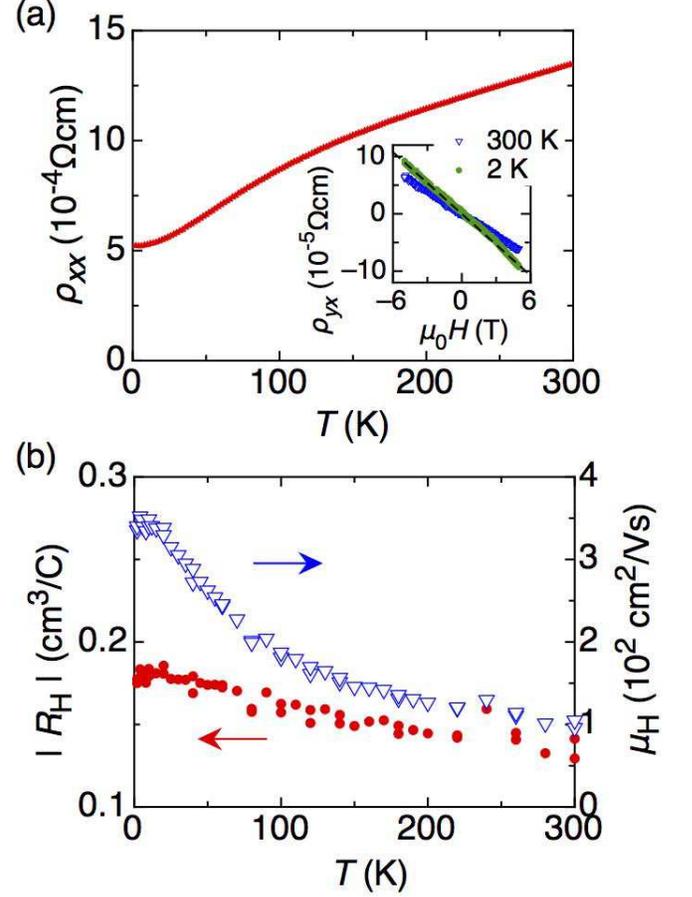}
\caption{(Color online) (a)~Temperature dependence of the longitudinal resistivity $\rho_{xx}$ in the range 2--300~K. The inset is the Hall resistivity $\rho_{yx}$ at 2 and 300~K. Negative magnetic field dependences indicate electron conduction. A linear fit of $\rho_{yx}$ at 2~K is also presented. (b)~Temperature dependence of the Hall coefficient $|R_{\rm{H}}|=|\rho_{yx}/\mu_0 H|$ and the Hall mobility $\mu_{\rm{H}}=|R_{\rm{H}}|/\rho_{xx}$. $|R_{\rm{H}}|$ at 2~K was $(1.775 \pm 0.006) \times10^{-1}$~cm$^3$/C and $\mu_{\rm{H}}$ at the same temperature was $342 \pm 6$~cm$^2$/Vs.}
\label{fig2}
\end{center}
\end{figure}

The temperature dependence of the longitudinal resistivity $\rho_{xx}$ is presented in Figure~\ref{fig2}(a). The inset shows the Hall resistivity $\rho_{yx}$ at 300 and 2~K. The positive temperature coefficient of $\rho_{xx}$ indicates metallic conduction. The negative magnetic coefficients of $\rho_{yx}$ indicate electron conduction. These results are consistent with the ARPES results; the Fermi level was located in the bulk conduction bands~\cite{PhysRevLett.105.136802,PhysRevLett.105.146801,PhysRevB.91.205306}. The $\rho_{yx}$ exhibited a linear magnetic field dependence at each temperature. The temperature dependence of the Hall coefficient $|R_{\rm{H}}|=|\rho_{yx}/B|$ and  the Hall mobility $\mu_{\rm{H}}=|R_{\rm{H}}|/\rho_{xx}$ are presented in Figure~\ref{fig2}(b). As shown in the figure, the $|R_{\rm{H}}|$ exhibited a small temperature dependence. $\mu_{\rm{H}}$ also exhibited a temperature variation above 15~K. $|R_{\rm{H}}|$ at 2~K was $(1.775 \pm 0.006) \times10^{-1}$~cm$^3$/C and $\mu_{\rm{H}}$ at the same temperature was $342 \pm 6$~cm$^2$/Vs. Because $\mu_{\rm{H}}$ is in proportion to the scattering time $\tau$, the temperature variation of $\mu_{\rm{H}}$ is attributed to the temperature dependence of $\tau$. We discuss the dependence later.

\begin{figure}[h]
\begin{center}
\includegraphics[width=\columnwidth,clip]{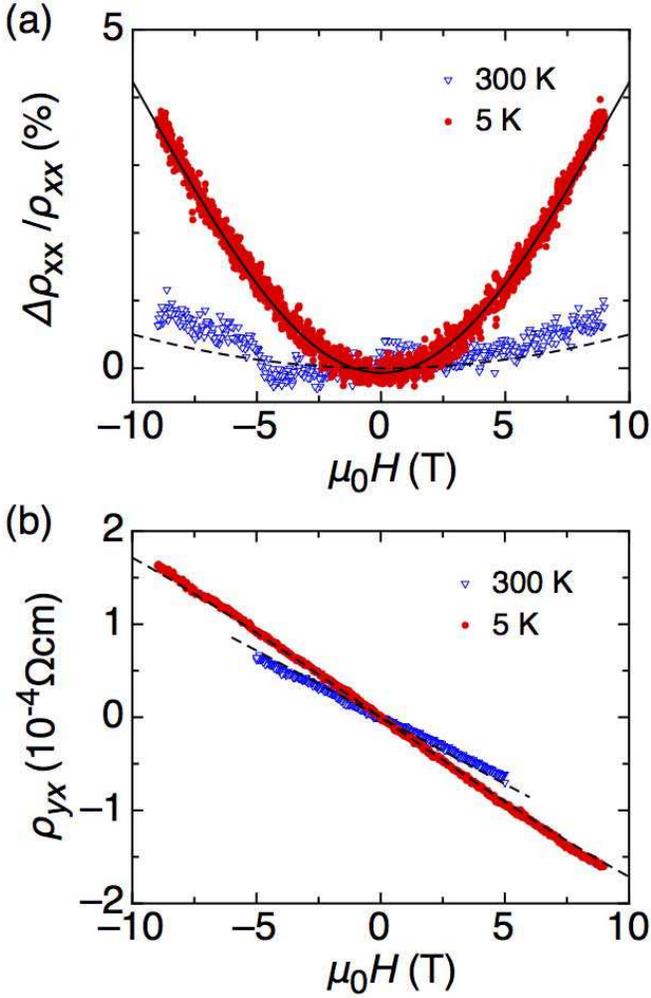}
\caption{(Color online) (a)~Magnetoresistance $\varDelta \rho_{xx} / \rho_{xx} = [\rho_{xx}(B) - \rho_{xx}(0)] / \rho_{xx}(B)$ at 300 and 5~K. The fit with Equation~(\ref{eq7}) at 5~K is shown by the solid line. The curve expected at 300~K is shown by the dashed line. (b) Hall resistivity $\rho_{yx}$ at 300 and 5~K. The curves expected at each temperature are shown by dashed lines.}
\label{fig3}
\end{center}
\end{figure}

$\rho_{xx}$ exhibited a magnetic field dependence as clearly seen in Figure~\ref{fig3}(a). Here $\varDelta \rho_{xx} / \rho_{xx} = [\rho_{xx}(B) - \rho_{xx}(0)] / \rho_{xx}(B)$ is the magnetoresistance. For further insight, we transformed the two-carrier model~\cite{Ashcroft_Mermin}. Within the model, $\rho_{xx}$ is expressed by Equation~(\ref{eq1}). Equation~(\ref{eq1}) was transformed to Equation~(\ref{eq2}) using $n_+=(n_1+n_2)$, $n_-=(n_1-n_2)$, $\mu_+=(\mu_1+\mu_2)/2$, $\mu_-=(\mu_1-\mu_2)/2$, and $N=n_-/n_+$, $M=\mu_-/\mu_+$. In the present case, the charges were $q_1=q_2=-e$, and thus, $-\mu_+>0$ and $-1 \leq M \leq 1$. We also defined $n_1 \geq n_2$; i.e., $0 \leq N \leq 1$. Finally, $\rho_{xx}(B)$, which was applied in further analyses, was obtained as Equation~(\ref{eq3}). The magnetoresistance was expressed as Equation~(\ref{eq4}).
\begin{widetext}
\begin{equation}
\rho_{xx}(B)= \frac{2}{n_+\mu_+}\Bigg[ \frac{q_1(1+N)(1+M) + q_2(1-N)(1-M) + [q_1(1+N)(1-M) + q_2(1-N)(1+M)](1-M^2)\mu_+^2B^2}{[q_1(1+N)(1+M)+q_2(1-N)(1-M)]^2 + [q_1(1+N)+q_2(1-N)]^2(1-M^2)^2\mu_+^2B^2} \Bigg]
\label{eq2}
\end{equation}
\begin{equation}
\rho_{xx}(B)= \frac{1}{n_+e(-\mu_+)}\Bigg[ \frac{1+NM + (1-M^2)(1-NM)\mu_+^2B^2}{(1+NM)^2+(1-M^2)^2\mu_+^2B^2} \Bigg]
\label{eq3}
\end{equation}
\begin{equation}
\frac{\varDelta\rho_{xx}}{\rho_{xx}}=  \frac{(1-N^2)M^2(1-M^2)\mu_+^2B^2}{(1+NM)^2+(1-N^2M^2)(1-M^2)\mu_+^2B^2}
\label{eq4}
\end{equation}
\end{widetext}
Within the model, $R_{\rm{H}}$ and $\mu_{\rm{H}}$ at $H=0$, which are presented in Figure~\ref{fig2}(b), are expressed as Equations~(\ref{eq5}) and~(\ref{eq6}). Using $\mu_{\rm{H}}$, Equation~(\ref{eq4}) was rewritten as Equation~(\ref{eq7}). Because $\mu_{\rm{H}}$ was an experimental value, $\varDelta\rho_{xx}/{\rho_{xx}}$ is determined employing two additional parameters, $N$ and $M$. Furthermore, $\rho_{yx}(B)$ was rewritten as Equation~(\ref{eq8}) using $R_{\rm{H}}$ and $\mu_{\rm{H}}$, and was also determined using $N$ and $M$.
\begin{widetext}
\begin{equation}
R_{\rm{H}} = -\frac{1}{n_+e} \Bigg[ \frac{1+2NM+M^2}{(1+NM)^2} \Bigg]
\label{eq5}
\end{equation}
\begin{equation}
\mu_{\rm{H}} = |\mu_+| \Bigg[ \frac{1+2NM+M^2}{1+NM} \Bigg]
\label{eq6}
\end{equation}
\begin{equation}
\frac{\varDelta\rho_{xx}}{\rho_{xx}} =  \frac{(1-N^2)M^2(1-M^2)\mu_{\rm{H}}^2B^2}{(1+2NM+M^2)^2+(1-N^2M^2)(1-M^2)\mu_{\rm{H}}^2B^2}
\label{eq7}
\end{equation}
\begin{equation}
{\rho_{yx}}(B)= R_{\rm{H}} \Bigg[ \frac{(1+2NM+M^2)^3B + (1+NM)^2(1-M^2)^2\mu_{\rm{H}}^2B^3}{(1+2NM+M^2)^3+(1+2NM+M^2)(1-M^2)^2\mu_{\rm{H}}^2B^2} \Bigg]
\label{eq8}
\end{equation}
\end{widetext}

$\varDelta \rho_{xx} / \rho_{xx}$ at 5~K and the fit with Equation~\ref{eq7} are presented in Figure~\ref{fig3}(a); the fit yielded $N=0.76\pm0.01$ and $M=-0.586 \pm 0.003$. $\varDelta \rho_{xx} / \rho_{xx}$ at 300~K, and the curve expected from the $N$ and $M$ values with $R_{\rm{H}}, \mu_{\rm{H}}$ at 300~K are also presented. The figure shows that the curve at 300~K is consistent with the experiments. $\rho_{yx}(B)$ and the curves expected from the same $N$ and $M$ values are presented in Figure~\ref{fig3}(b). The curves also agree well with the experimental results. We thus conclude that the $N$ and $M$ values obtained are invariant within the studied temperature range, and the transport properties of the sample are explained well with the two-channel model.

Once the four parameters of the two-channel model $R_{\rm{H}}$, $\mu_{\rm{H}}$, $N$, and $M$ were obtained, the densities and mobilities of high- and low-mobility electrons were determined. The densities and the mobilities at 5~K are summarized in Table~\ref{tab1}. Here $n_1=n_+(1+N)/2$, $n_2=n_+(1-N)/2$, $\mu_1=\mu_+(1+M)$, and $\mu_2=\mu_+(1-M)$. 

\begin{table}[h]
\caption{Electric transport properties at 5~K obtained from the two-carrier model. $n_1$ and $n_2$, and $\mu_1$ and $\mu_2$ are respectively the densities and mobilities of the electrons.}
\label{tab1}
\begin{ruledtabular}
	\begin{tabular}{ccccc} %\hline
		& $n_1$ & $4.45 \pm 0.01$ & $10^{19}$~/cm$^3$ & \\
		& $n_2$ & $0.06 \pm 0.01$ & $10^{19}$~/cm$^3$  & \\
\\
		& $\mu_1$ & $176 \pm 2$ & cm$^2$/Vs & \\
		& $\mu_2$ & $676 \pm 2$ & cm$^2$/Vs  & \\ %\hline
	\end{tabular}
\end{ruledtabular}
\end{table}

\section{Discussion}
\begin{figure}[h]
\begin{center}
\includegraphics[width=\columnwidth,clip]{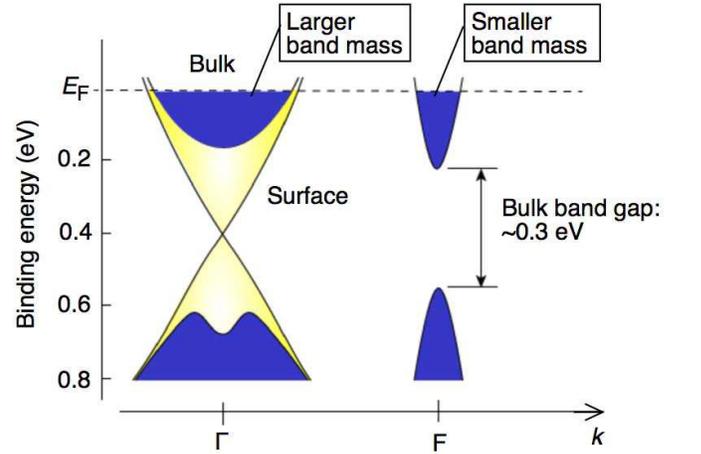}
\caption{(Color online) A schematic of the energy band dispersion of TlBiSe$_2$ revealed by ARPES~\cite{PhysRevLett.105.146801,PhysRevB.91.205306,comment2}. The smaller band curvature around the $\Gamma$ point results in a larger band mass of electrons. The greater band curvature around the $F$ point results in smaller band mass. The amplitude of the minimum bulk band gap is also presented.}
\label{fig4}
\end{center}
\end{figure}

As demonstrated above, the magnetotransport properties were well explained by the two-channel model. We next discuss the origin of these conduction channels with the electronic band structure. A schematic of the structure revealed by ARPES is presented in Figure~\ref{fig4}~\cite{PhysRevLett.105.146801,PhysRevB.91.205306,comment2}. The figure shows surface Dirac fermions and two Fermi surfaces of the bulk; one of the Fermi surfaces was around the $\Gamma$ point and the other around the $F$ point. For the surface Dirac fermions, the two-dimensional carrier density $n_{\rm{s}}=k_{\rm{F}}^2/(4\pi)$ was estimated as $\sim 8 \times10^{12}$~/cm$^2$, where $k_{\rm{F}}$ was the Fermi wave number obtained from the ARPES spectrum (Figure 2 of Ref.~\cite{PhysRevLett.105.146801}). The value of $n_{\rm{s}}$ yielded an effective 3D carrier density $n_{\rm{s}}/d = 2.1 \times10^{14}$~/cm$^3$, where $d=0.380$~mm was the sample thickness. The estimated value of $n_{\rm{s}}/d$ was smaller than the experimental values by five orders of magnitude. This indicates that the contribution of the Dirac fermions to $\rho_{xx}$ and $\rho_{yx}$ was negligibly small; i.e., the transport properties are attributed to the bulk electrons. Figure~\ref{fig4} shows that the Fermi surface around the $\Gamma$ point has a larger Fermi cross section and a larger band mass than that around the $F$ point. These band characters are consistent with the two-carrier model, where the Fermi surface around the $\Gamma$ point yields higher carrier density and lower mobility; i.e., higher $n_1$ and lower $\mu_1$. It is also concluded that the minor electrons with carrier density $n_2$, and mobility $\mu_2$ can be attributed to the Fermi surface around the $F$ point. 

As discussed above, novel surface transport is hidden in bulk metallic conduction with carrier density $\sim 10^{19}$~/cm$^3$. In contrast, bulk insulating state with carrier density $< 10^{17}$~/cm$^3$ was demonstrated for Tl$_{1-x}$Bi$_{1+x}$Se$_{2-\delta}$ ($x \leq 0.028, \delta=0.28$) and the surface metallic conduction was distinguished~\cite{PhysRevB.91.205306,PhysRevB.90.201307}. These results indicate a presence of metal-insulator transition in bulk between $x=0.064$ and $0.028$, and novel surface transport can be distinguished only in the bulk insulating samples.

\begin{figure}[h]
\begin{center}
\includegraphics[width=\columnwidth,clip]{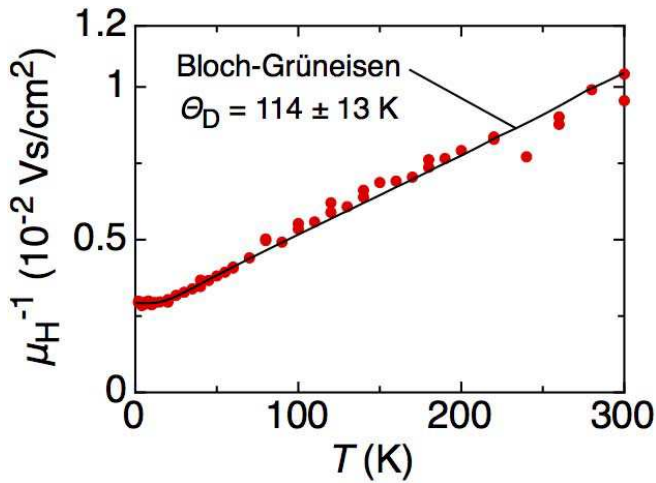}
\caption{(Color online) Temperature dependence of the inverse Hall mobility $\mu_{\rm{H}}^{-1}$. The fit with the Bloch-Gr{\"u}neisen formula yields the Debye temperature $\varTheta_{\rm{D}}=114 \pm 13$~K. See the text for details.}
\label{fig5}
\end{center}
\end{figure}

Finally, we discuss the temperature dependence of scattering realized in the system. The scattering time is in proportion to the mobility, and the temperature dependence of $\mu_+$ is thus attributed to the temperature dependence of average scattering time $\tau$. Note that the agreement with the two-carrier model indicates that the scattering was insensitive to the magnetic field. As discussed above, the temperature dependences of $N$ and $M$ were negligibly small. These results indicate that $\mu_{\rm{H}}$ can be expressed as $\mu_{\rm{H}}=0.816|\mu_+|$ using the $N$ and $M$ values obtained (see Equation~\ref{eq6}) and that the temperature dependence of $\mu_{\rm{H}}$ can be attributed to $\tau$. The temperature dependence of $\mu_{\rm{H}}^{-1}$ is presented in Figure~\ref{fig5}. The fit with the Bloch-Gr{\"u}neisen formula
\begin{equation}
\frac{1}{\tau} \propto \Bigg(\frac{T^5}{\varTheta_{\rm{D}}^6}\Bigg)\int_0^{\frac{\varTheta_{\rm{D}}}{T}}\frac{x^5}{(e^x-1)(1-e^{-x})} dx,
\label{eq9}
\end{equation}
which describes the temperature dependence of acoustic phonon scattering, is also presented. Here $\varTheta_{\rm{D}}$ is the Debye temperature determined in the present transport experiment. As shown in Figure~\ref{fig5}, the temperature dependence of $1/\mu_{\rm{H}} \propto 1/\tau$ agrees well with the formula, which yields $\varTheta_{\rm{D}}=113 \pm 14$~K. Note that $\varTheta_{\rm{D}}$ agrees with the Debye temperature determined from the specific heat~\cite{TlBiSe2debye}. The result indicates that the scattering was dominated by acoustic phonons. Similar results are often seen for metals of a weakly correlated electron system, indicating that the scattering of bulk electrons is explained by the conventional theory of metals. 

\section{Conclusion}
We investigated the transport properties of the 3D topological insulator TlBiSe$_2$. Two-carrier transport properties were precisely determined by applying a newly developed analysis procedure. The magnetotransport properties were well explained accounting for high- and low-mobility electrons in the whole temperature range. This indicated that the scattering of electrons was insensitive to the magnetic field. The transport properties obtained were consistent with ARPES results, and contributions of the surface transport were negligibly small over the entire temperature range studied. The scattering of bulk electrons was dominated by acoustic phonons. The results indicate that the bulk transport properties of the TlBiSe$_2$ sample were in the vicinity of those of conventional metals. These simple nature of bulk conduction hints proper verification of novel surface transport properties. It should be emphasized that the analysis using conventional two-carrier model with four active parameters leads to large errors. Instead, the analysis proposed in this study exhibited sufficiently small errors, and therefore possessed statistical reliability. We also note that the procedure should be useful for all other transport studies.

\section{Acknowledgements}
This work was partly supported by JSPS KAKENHI, a Grant-in-Aid for JSPS Scientific Research (B) (Grant No. 23340105), and Innovative Areas "Nano Spin Conversion Science". G. E. and K. K. acknowledge support from from the Japan Society for the Promotion of Science for Young Scientists.

\end{document}